\documentclass{rnaastex}
\usepackage{comment, todonotes}

\newcommand{\gbr}{\ensuremath{(G_{\rm BP} - G_{\rm RP})}}
\def\amin{\ifmmode^{\prime}\else$^{\prime}$\fi}
\newcommand{\Msun}{\ifmmode {M_{\odot}}\else{M$_{\odot}$}\fi}
\newcommand{\lapprox }{{\lower0.8ex\hbox{$\buildrel <\over\sim$}}}
\newcommand{\gapprox }{{\lower0.8ex\hbox{$\buildrel >\over\sim$}}}
\def\Prot{$P_{\mathrm{rot}}$}

\shorttitle{Re-crowning The Queen}
\shortauthors{Singh et al.}

\begin{document}

\title{Re-crowning The Queen: Membership, Age and Rotation Periods for \\ the Open Cluster Coma Berenices}

\newcommand{\amnh}{American Museum of Natural History, 200 Central Park West, New York, NY 10024, USA}
\newcommand{\columbia}{Department of Astronomy, Columbia University, 550 West 120th Street, New York, NY 10027, USA}

\author{Kyle Singh}
\affiliation{\columbia}

\author{Peter Rothstein}
\affiliation{\columbia}

\author[0000-0002-2792-134X]{Jason L.~Curtis}
\affiliation{\columbia}
\affiliation{\amnh}

\author[0000-0002-8047-1982]{Alejandro N\'u\~nez}
\affiliation{\columbia}

\author[0000-0001-7077-3664]{Marcel A.~Ag\"{u}eros}
\affiliation{\columbia}

\begin{abstract}
Coma Berenices (Coma Ber), an open cluster about the same age as Praesepe and the Hyades (700-800 Myr) is, despite being only 85 pc away, less well studied than its famous cousins. This is due principally to its sparseness and low proper motion, which together made Coma Ber's membership challenging to establish pre-Gaia. We have curated a new list of its members based on Gaia DR2 astrometry, derived its metallicity and interstellar reddening using LAMOST data, and inferred the cluster's age by fitting PARSEC isochrones to its color--magnitude diagram. We then measured rotation periods for Coma Ber's low-mass members using TESS and ZTF photometry. Our isochrone fitting and the TESS- and ZTF-derived rotation periods confirm that Coma Ber is coeval with the Hyades and Praesepe. This work is the first step toward re-establishing Coma Ber as another valuable benchmark cluster for age--rotation--activity studies.
\end{abstract}

\section{Introduction} \label{sec:intro}
A low-mass ($\lapprox$1 \Msun) star's angular-momentum content  depends on its internal structure, core--envelope coupling, magnetic field geometry, and tidal effects due to any companions. Determining the distribution of rotation periods (\Prot) for these stars in open clusters, whose ages are well known, provides empirical constraints on how the angular-momentum content changes over time and as a function of mass, informing theoretical models. 
However, there are only a limited number of nearby open clusters whose low-mass members can be examined in detail. Studies of the age--rotation relation have generally relied on observations of rich, close-by clusters such as the Hyades, Pleiades, and Praesepe. With new observing capabilities, however, this work has been extended in the last few years to more distant and challenging clusters.

We report on our work to add Coma Berenices (Coma Ber) to the small number of benchmark open clusters for age--rotation--activity studies. Coma Ber, which is about the same age as  Praesepe ($\approx$700 Myr) and, at 87 pc, not much farther away than the Hyades, has been far less well studied than its famous cousins, mainly because its sparseness and low proper motion made its membership challenging to establish. With Gaia data, however, it is now possible to assemble a high-confidence membership catalog for this cluster. Below we describe how we constructed such a catalog, used it to find an isochrone age for the cluster, and then determined the color--period distribution for low-mass stars in Coma Ber based on light curves from the Transiting Exoplanet Survey Satellite \citep[TESS;][]{ricker} 
and the Zwicky Transient Factory \citep[ZTF;][]{masci2018}.

Our isochrone fitting and the TESS- and ZTF-derived \Prot\ confirm that Coma Ber is coeval with the Hyades and Praesepe, making Coma Ber another valuable benchmark for studies of the age--rotation relation.

\begin{figure}[t!]
    \centering
    \includegraphics[width=6.5in]{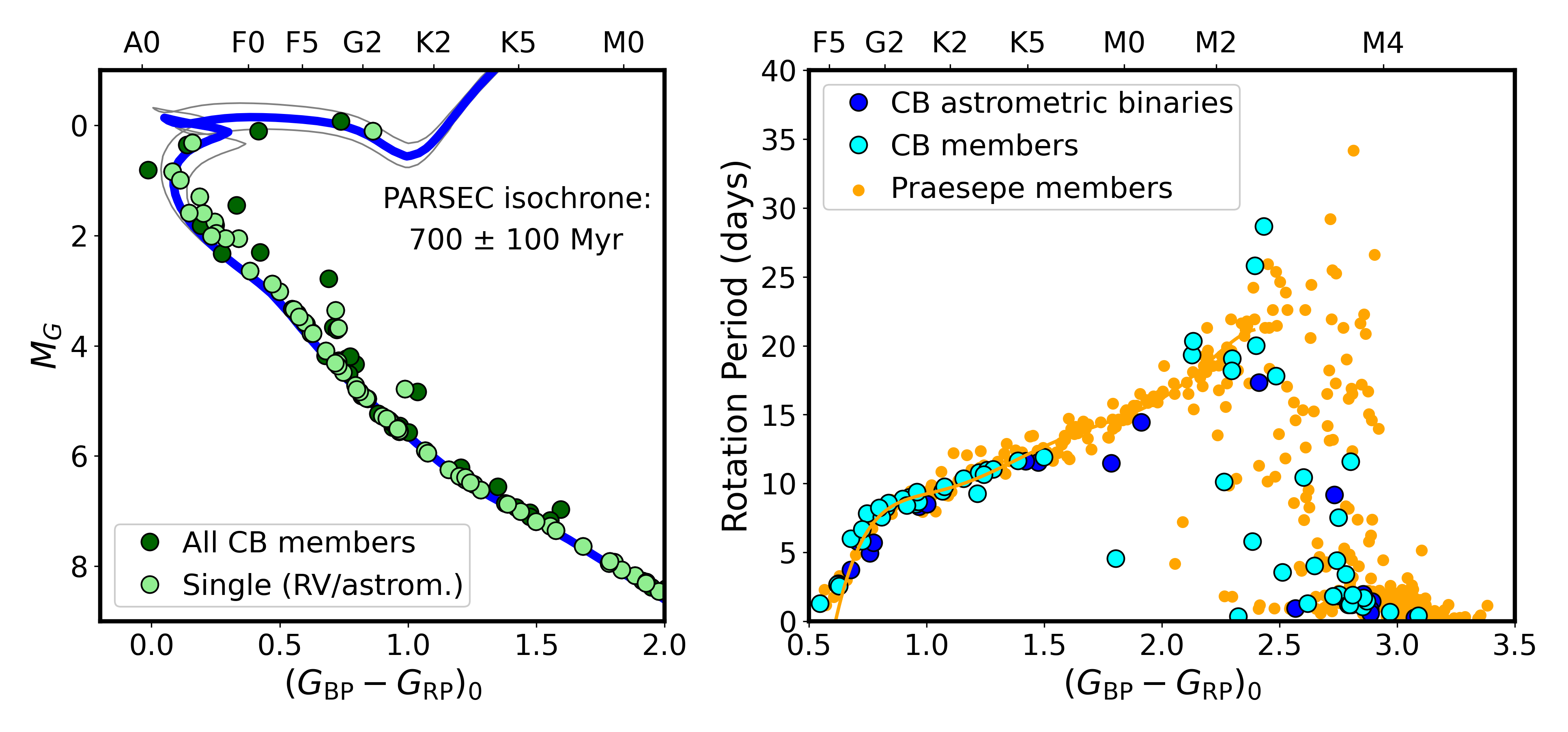}
    \caption{{\it Left---}Gaia EDR3 CMD for Coma Ber members (291 stars) with best-fit PARSEC isochrone overlaid. The subset of cluster members that have Gaia DR2 RVs and RV errors and EDR3 RUWEs and proper motions consistent with being single stars are highlighted in light green (208 stars). {\it Right---}Distribution of \Prot\ for stars consistent with being single members of Coma Ber 
    (62 stars; teal points) together with data for single-star members of Praesepe \citep[orange points;][]{douglas2019}; the orange line is a polynomial fit to the slowly rotating sequence in Praesepe. Dark blue points show Coma Ber astrometric binary candidates that are photometrically and spectroscopically single (i.e., there no evidence that they have short orbital periods where tides operate, nor is the secondary  bright enough to impact the time-series analysis; 19 stars). The slowly rotating sequences of FGK stars for the two clusters are remarkably similar, despite the 0.2 dex difference in metallicity between them.}
    \label{fig:cpd}
\end{figure}

\section{A new membership catalog} \label{sec:members}

We combined the pre-Gaia list from \citet{kraus2007} with Gaia-DR2-based lists from \citet{DR2HRD}, \citet{FMA2019}, \citet{Tang2019}, and results of our own query of the DR2 database. We later updated the astrometric and photometric values for our candidate members to their Early Data Release 3 \citep[EDR3;][]{edr3} values. 

By plotting the EDR3 parallaxes for these stars, we found that $\approx$10\% of our candidate members have very low parallaxes, indicating that they are likely background giants. 
We also used Gaia data to identify 
binary candidates following criteria based on those used in \citet{curtis2020}. Our new catalog will be described fully in a forthcoming paper (Ag\"ueros et al.~in prep.). 

\section{Coma Ber's Fundamental Properties} \label{sec:feh}

We collected spectroscopic properties for Coma Ber and Praesepe members from \citet{Xiang2019}, which is based on the Large Sky Area Multi-Object Fiber Spectroscopic Telescope (LAMOST) DR5. Using Praesepe as a reference, we fit and removed a systematic trend in [Fe/H] vs.~effective temperature, thus placing the values for Coma Ber on a differential scale, and found that the metallicity of Coma Ber is 0.21 dex lower than Praesepe. Applying [Fe/H] = +0.15 dex for Praesepe, as determined from Keck/HIRES spectra, we find [Fe/H]~=~$-$0.06~dex for Coma Ber.

We computed the interstellar reddening for Coma Ber by comparing spectroscopic temperatures from LAMOST with photometric \gbr\ colors from Gaia. Here again, we obtain a relative value by calculating the reddening relative to that for Praesepe. 
Applying the visual extinction value of $A_V = 0.04$ mag for Praesepe \citep{douglas2019}, we find zero reddening in the Coma Ber foreground.

To determine the age of Coma Ber, we 
constructed EDR3 color-magnitude diagrams (CMDs) for it and for Praesepe. 
We then used PAdova and TRieste Stellar Evolution Code (PARSEC) isochrones \citep{parsec} to determine the ages for both clusters.
We visually fit the main sequence turnoff and subgiant members. The best fit for Coma Ber is a 700 Myr isochrone (see left panel, Figure~\ref{fig:cpd}). For Praesepe, the best fit is 730 Myr, an age that is slightly older than the median literature age (670 Myr) for this cluster. We conclude that the clusters are coeval.

\section{Rotation In Coma Ber} \label{sec:prot}

TESS surveyed members of Coma Ber in Sectors 15 or 16. Given the relatively short sector duration ($\approx$27 days) and large data gaps midway through each sector during the data downlink, our analysis was restricted to searching for \Prot~$\leq$~10 days. 

ZTF has observed Coma Ber regularly since 2018 March 25. Following \citet{curtis2020}, we extracted differential photometry from the archival ZTF imaging. Although the resulting light curves are much sparser and more irregularly sampled than those from TESS, they can  provide confirmation of the TESS-derived periods in cases where a star was observed from the ground and from space, and can be used to measure periods for slower rotators in Coma~Ber. 

We merged our TESS (58 stars) and ZTF measurements (26 stars), with data from \citet[][15 stars]{collier2009} and \citet[][4 stars]{terrien2014}, yielding a total of 81 
Coma Ber stars with measured \Prot. We then filtered out stars that are spectroscopic and/or photometric binaries
\footnote{For example, we removed stars with Gaia renormalized unit weight error (RUWE) values $>$1.2, which are likely to be in wide binaries.} 
and stars with bright members blended in the large TESS pixels. The resulting color--period  distribution for Coma Ber is shown in the right panel of Figure~\ref{fig:cpd}, where we also include the distribution of single-star rotators in Praesepe. 
The slowly rotating sequences of FGK stars for the two clusters are remarkably similar, despite the 0.2 dex difference in metallicity between them.


\begin{acknowledgments}
We thank John Brewer for sharing spectroscopic results for Hyades, Praesepe, and Coma Ber, and Soichiro Hattori for assistance with \texttt{tess\_cpm.py}. 

\end{acknowledgments}




\begin{thebibliography}{}
\expandafter\ifx\csname natexlab\endcsname\relax\def\natexlab#1{#1}\fi
\providecommand{\url}[1]{\href{#1}{#1}}
\providecommand{\dodoi}[1]{doi:~\href{http://doi.org/#1}{\nolinkurl{#1}}}
\providecommand{\doeprint}[1]{\href{http://ascl.net/#1}{\nolinkurl{http://ascl.net/#1}}}
\providecommand{\doarXiv}[1]{\href{https://arxiv.org/abs/#1}{\nolinkurl{https://arxiv.org/abs/#1}}}

\bibitem[{{Bressan} {et~al.}(2012){Bressan}, {Marigo}, {Girardi}, {Salasnich},
  {Dal Cero}, {Rubele}, \& {Nanni}}]{parsec}
{Bressan}, A., {Marigo}, P., {Girardi}, L., {et~al.} 2012, \mnras, 427, 127,
  \dodoi{10.1111/j.1365-2966.2012.21948.x}

\bibitem[{Collier~Cameron {et~al.}(2009)Collier~Cameron, Davidson, Hebb,
  Skinner, Anderson, Christian, Clarkson, Enoch, Irwin, Joshi, Haswell,
  Hellier, Horne, Kane, Lister, Maxted, Norton, Parley, Pollacco, Ryans,
  Scholz, Skillen, Smalley, Street, West, Wilson, \& Wheatley}]{collier2009}
Collier~Cameron, A., Davidson, V.~A., Hebb, L., {et~al.} 2009, Monthly Notices
  of the Royal Astronomical Society, 400, 451,
  \dodoi{10.1111/j.1365-2966.2009.15476.x}

\bibitem[{Curtis {et~al.}(2020)Curtis, Agüeros, Matt, Covey, Douglas, Angus,
  Saar, Cody, Vanderburg, Law, Kraus, Latham, Baranec, Riddle, Ziegler, Lund,
  Torres, Meibom, Aguirre, \& Wright}]{curtis2020}
Curtis, J.~L., Agüeros, M.~A., Matt, S.~P., {et~al.} 2020, The Astrophysical
  Journal, 904, 140, \dodoi{10.3847/1538-4357/abbf58}

\bibitem[{{Douglas} {et~al.}(2019){Douglas}, {Curtis}, {Ag{\"u}eros},
  {Cargile}, {Brewer}, {Meibom}, \& {Jansen}}]{douglas2019}
{Douglas}, S.~T., {Curtis}, J.~L., {Ag{\"u}eros}, M.~A., {et~al.} 2019,
  Astrophysical Journal, 879, 100, \dodoi{10.3847/1538-4357/ab2468}

\bibitem[{{F{\"u}rnkranz} {et~al.}(2019){F{\"u}rnkranz}, {Meingast}, \&
  {Alves}}]{FMA2019}
{F{\"u}rnkranz}, V., {Meingast}, S., \& {Alves}, J. 2019, \aap, 624, L11,
  \dodoi{10.1051/0004-6361/201935293}

\bibitem[{{Gaia Collaboration} {et~al.}(2020){Gaia Collaboration}, {Brown},
  {Vallenari}, {Prusti}, {de Bruijne}, {Babusiaux}, \& {Biermann}}]{edr3}
{Gaia Collaboration}, {Brown}, A.~G.~A., {Vallenari}, A., {et~al.} 2020, arXiv
  e-prints, arXiv:2012.01533.
\newblock \doarXiv{2012.01533}

\bibitem[{{Gaia Collaboration} {et~al.}(2018){Gaia Collaboration}, {Babusiaux},
  {van Leeuwen}, {Barstow}, {Jordi}, {Vallenari}, {Bossini}, {Bressan},
  {Cantat-Gaudin}, {van Leeuwen}, \& et~al.}]{DR2HRD}
{Gaia Collaboration}, {Babusiaux}, C., {van Leeuwen}, F., {et~al.} 2018,
  Astronomy \& Astrophysics, 616, A10, \dodoi{10.1051/0004-6361/201832843}

\bibitem[{{Kraus} \& {Hillenbrand}(2007)}]{kraus2007}
{Kraus}, A.~L., \& {Hillenbrand}, L.~A. 2007, Astronomical Journal, 134, 2340,
  \dodoi{10.1086/522831}

\bibitem[{Masci {et~al.}(2018)Masci, Laher, Rusholme, Shupe, Groom, Surace,
  Jackson, Monkewitz, Beck, Flynn, Terek, Landry, Hacopians, Desai, Howell,
  Brooke, Imel, Wachter, Ye, Lin, Cenko, Cunningham, Rebbapragada, Bue, Miller,
  Mahabal, Bellm, Patterson, Juri{\'{c}}, Golkhou, Ofek, Walters, Graham,
  Kasliwal, Dekany, Kupfer, Burdge, Cannella, Barlow, Sistine, Giomi, Fremling,
  Blagorodnova, Levitan, Riddle, Smith, Helou, Prince, \& Kulkarni}]{masci2018}
Masci, F.~J., Laher, R.~R., Rusholme, B., {et~al.} 2018, Publications of the
  Astronomical Society of the Pacific, 131, 018003,
  \dodoi{10.1088/1538-3873/aae8ac}

\bibitem[{{Ricker} {et~al.}(2015){Ricker}, {Winn}, {Vanderspek}, {Latham},
  {Bakos}, {Bean}, {Berta-Thompson}, {Brown}, {Buchhave}, {Butler}, {Butler},
  {Chaplin}, {Charbonneau}, {Christensen-Dalsgaard}, {Clampin}, {Deming},
  {Doty}, {De Lee}, {Dressing}, {Dunham}, {Endl}, {Fressin}, {Ge}, {Henning},
  {Holman}, {Howard}, {Ida}, {Jenkins}, {Jernigan}, {Johnson}, {Kaltenegger},
  {Kawai}, {Kjeldsen}, {Laughlin}, {Levine}, {Lin}, {Lissauer}, {MacQueen},
  {Marcy}, {McCullough}, {Morton}, {Narita}, {Paegert}, {Palle}, {Pepe},
  {Pepper}, {Quirrenbach}, {Rinehart}, {Sasselov}, {Sato}, {Seager},
  {Sozzetti}, {Stassun}, {Sullivan}, {Szentgyorgyi}, {Torres}, {Udry}, \&
  {Villasenor}}]{ricker}
{Ricker}, G.~R., {Winn}, J.~N., {Vanderspek}, R., {et~al.} 2015, Journal of
  Astronomical Telescopes, Instruments, and Systems, 1, 014003,
  \dodoi{10.1117/1.JATIS.1.1.014003}

\bibitem[{{Tang} {et~al.}(2019){Tang}, {Pang}, {Yuan}, {Chen}, {Hong},
  {Goldman}, {Just}, {Shukirgaliyev}, \& {Lin}}]{Tang2019}
{Tang}, S.-Y., {Pang}, X., {Yuan}, Z., {et~al.} 2019, \apj, 877, 12,
  \dodoi{10.3847/1538-4357/ab13b0}

\bibitem[{Terrien {et~al.}(2014)Terrien, Mahadevan, Deshpande, Bender, Cargile,
  Hearty, Cottaar, Allende~Prieto, Fleming, Frinchaboy, Jackson, Johnson,
  Majewski, Nidever, Pepper, Rodriguez, Schneider, Siverd, Stassun, Weaver, \&
  Wilson}]{terrien2014}
Terrien, R.~C., Mahadevan, S., Deshpande, R., {et~al.} 2014, Astrophysical
  Journal, 782, 61, \dodoi{10.1088/0004-637X/782/2/61}

\bibitem[{Xiang {et~al.}(2019)Xiang, Ting, Rix, Sandford, Buder, Lind, Liu,
  Shi, \& Zhang}]{Xiang2019}
Xiang, M., Ting, Y.-S., Rix, H.-W., {et~al.} 2019, The Astrophysical Journal
  Supplement Series, 245, 34, \dodoi{10.3847/1538-4365/ab5364}

\end{thebibliography}
\bibliographystyle{aasjournal}

\end{document}